\documentclass{svproc}
\usepackage{url}

\usepackage{graphicx}
\usepackage{xspace}
\usepackage{amsmath}
\usepackage{slashed}
\usepackage[numbers,sort&compress]{natbib}

\newcommand{\ket}[1]{\ensuremath{\left|#1\right\rangle}\xspace}
\newcommand{\bra}[1]{\ensuremath{\left\langle #1\right|}\xspace}
\newcommand{\g}{\ensuremath{\gamma}\xspace}
\newcommand{\G}[0]{\ensuremath{\Gamma}}
\newcommand{\da}{\ensuremath{\varphi(x_i)}}

\begin{document}
\mainmatter              
\title{Nucleon Parton Distribution Amplitude:\\ A scalar diquark picture}
\titlerunning{Nucleon PDA}  
%
\author{C\'edric Mezrag\inst{1}, Jorge Segovia\inst{2}, Minghui Ding\inst{3,4,5},\\ Lei Chang\inst{4} and Craig D. Roberts\inst{3}}
\authorrunning{C\'edric Mezrag \emph{et al.}} 
%
\tocauthor{C\'edric Mezrag, Jorge Segovia, Minghui Ding, Lei Chang, Craig D. Roberts}
\institute{Istituto  Nazionale  di  Fisica  Nucleare,  Sezione  di  Roma,\\  P. le  A. Moro  2,  I-00185  Roma,  Italy,\\
\email{cedric.mezrag@roma1.infn.it},
\and
Departamento de Sistemas F\'isicos, Qu\'imicos y Naturales,\\
Universidad Pablo de Olavide, E-41013 Sevilla, Spain
\and
Physics Division, Argonne National Laboratory, Argonne, Illinois 60439, USA
\and
School  of  Physics,  Nankai  University,  Tianjin  300071,  China
\and
European Centre for Theoretical Studies in Nuclear Physics\\ and Related Areas (ECT$^\ast$) and Fondazione Bruno Kessler,\\
Villa Tambosi, Strada delle Tabarelle 286, I-38123 Villazzano (TN) Italy
}

\maketitle              

\begin{abstract}
We report progress on the development of Perturbative Integral Representation Ans\"atze to compute the nucleon Faddeev Wave function, using an explicit quark-diquark picture. Our formalism is able to handle non-pointlike diquark and to mimic the strong, dynamical correlations observed when solving the Faddeev Equation. We then project the wave function in order to compute the leading-twist Parton Distribution Amplitude.

\end{abstract}
\section{Introduction}
One of the main objective of modern physics is to manage to describe hadron structure in terms of quarks and gluons, the fundamental degrees of freedom of Quantum Chromodynamics (QCD). A major effort is undertaken, both on the experimental and theoretical sides. Insight of hadron structure can be gained experimentally thanks to the so-called factorisation theorem, which allows us to split cross sections or amplitudes of various inclusive or exclusive processes into a ``hard part'', expandable in perturbation theory, and ``soft part'' encoding the non-perturbative information on hadron structure. Within this framework, parton distribution amplitudes (PDAs) play an important role in our understanding of exclusive processes at large momentum transfer \cite{Collins:1989gx,Efremov:1979qk,Lepage:1980fj,Chernyak:1984bm}.

The PDAs can be computed from the Lightfront Wave Functions (LFWFs) by integrating out, to a certain scale $\zeta$, the transverse momentum degrees of freedom of every parton belonging to the considered Fock state. If the energy involved in the scattering event is high enough, the so-called leading twist PDA, coming from one of the projection of the lowest Fock state is expected to dominate the description of exclusive processes. Contrary to the light-quark meson sector, where a significant effort has been performed in the recent years both using continuum techniques \cite{Chang:2013pq,Gao:2014bca,Shi:2015esa} and lattice-QCD \cite{Segovia:2013eca,Braun:2015axa,Bali:2017ude,Zhang:2017bzy}, the nucleon leading-twist PDA remains to be computed at experimentally available energy, as only few modern attempts have partially tackled the issue  \cite{Braun:2014wpa,Bali:2015ykx,Mezrag:2017znp}. However, its limit when the typical scale, $\zeta$, goes to infinity is known and called the asymptotic PDA.

In this paper, we report the results (and the improvements made since \cite{Mezrag:2017znp}) of our approach to compute an insightful PDA for the nucleon. It relies on two main ingredients. First, we take advantage of the emerging picture coming from decades of work on solving the Faddeev equation \cite{Cahill:1988dx,Burden:1988dt,Cahill:1988zi,Reinhardt:1989rw,Efimov:1990uz} yielding a borromean picture of the nucleon \cite{Segovia:2015ufa}. This picture, which implies the constant breaking and formation of dynamical diquark correlations within baryons, has scored phenomenological successes (see \emph{e.g.} \cite{Segovia:2013uga,Segovia:2013rca,Segovia:2014aza,Segovia:2015hra,Segovia:2016zyc}). Then, the Perturbation-Theory Integral Representation (PTIR) introduced by Nakanishi \cite{Nakanishi:1963zz,Nakanishi:1969ph} plays the second essential role in our approach. It allows us to write Bethe-Salpeter and Faddeev amplitudes in terms of a known momentum-dependent kernel and a momentum-independent Nakanishi weight function. Interestingly, PTIR is proved to be valid at all order of perturbation theory, and has already been used successfully in the meson sector to compute PDAs and beyond, Parton Distribution Functions (PDFs) and Generalised Parton Distributions (GPDs) (see \cite{Chang:2013pq,Frederico:2011ws,Carbonell:2017kqa,dePaula:2017ikc,Salme:2017oge,Chang:2014lva,Mezrag:2014jka,Mezrag:2014tva,Mezrag:2016hnp,Chouika:2017dhe,Chouika:2017rzs,Shi:2018mcb}).

\section{Generalities}
In order to define the nucleon PDA, we need to introduce two light-like vectors $n$ and $p$, related to the nucleon momentum $P$ such that $ p^\mu  =  P^\mu - n^\mu P^2/(2 P\cdot n)$. 
Introducing a three-quarks matrix element, we can define the nucleon leading twist PDA in Euclidean space as:
\begin{align}
  \label{eq:PhiasMatrix1}
  \bra{0}\epsilon^{ijk}&\left(\tilde{u}_\uparrow^{i}(z_1 n)C^\dagger\slashed{n}u_\downarrow^{j}(z_2 n)\right)\slashed{n}d_\uparrow^{k}(z_3 n)\ket{P,\lambda} \nonumber \\
  &= i\frac{1}{2} (p \cdot n)f_N \slashed{n} B^\uparrow\int \mathcal{D}x \varphi([x]) e^{-ip \cdot n\sum_i x_i z_i},
\end{align}
where $\tilde{u}$ is the quark field $u$ transposed, $B$ is the baryon Euclidean Dirac spinor, $f_N$ is the normalisation constant of the PDA, or the value of the wave function at the origin, $[x] = (x_1,x_2,x_3)$, and
\begin{equation}
  \label{eq:Conventions}
  \mathcal{D}x = \textrm{d}x_1\textrm{d}x_2\textrm{d}x_3\delta\left(1-\sum_{i=1}^3x_i \right) \quad \textrm{and} \quad q_{\uparrow\downarrow} = L^{\uparrow\downarrow}q =\frac{1\pm \gamma_5}{2}q.
\end{equation}
The PDA can be readily computed once the nucleon Faddeev wave function $\chi$ is known. The latter is the non-amputated version of the Faddeev amplitude $\Psi$. As already emphasised in the introduction, modern studies of the Faddeev amplitude strongly suggest the existence of dynamical diquark correlations\footnote{We stress that these diquarks are \emph{not} the elementary ones introduced 50 years ago.}, and we therefore describe the amplitude as:
\begin{equation}
  \label{eq:FaddeevAmplitude}
  \Psi = \psi_1 +\psi_2+ \psi_3,
\end{equation}
where the labels refer to the quark bystander. $\psi_{1,2}$ can be deduced from $\psi_3$ by cyclic permutations of the indices. One can decompose:
\begin{eqnarray}
  \label{eq:Psi3Decomposition}
  \psi_3 & = & \mathcal{N}_3^0 + \mathcal{N}_3^1\\
  \label{eq:Psi3Scalar}
  \mathcal{N}_3^0  & = & \left[\G^0(k,K)\right]^{\alpha_1 \alpha_2}_{\tau_1 \tau_2} \Delta^0(K)  \left[\mathcal{S}(\ell;P) B(P) \right]^{\alpha_3}_{\tau_3}, \\
  \label{eq:Psi3AV}
  \mathcal{N}_3^1  & = & \left[\G^{j;1}_{\mu}(k,K)\right]^{\alpha_1 \alpha_2}_{\tau_1 \tau_2} \Delta^{1}_{\mu\nu}(K)  \left[\mathcal{A}^j_\nu(\ell;P) B(P) \right]^{\alpha_3}_{\tau_3}
\end{eqnarray}
where $\mathcal{N}_3^0$ and $\mathcal{N}_3^1$ are respectively the scalar and axial-vector diquark contributions to $\psi_3$, $(\{p\},\{\alpha\},\{\sigma\})$ are respectively the momenta, Dirac and isospin labels of the nucleon Faddeev Amplitude. We have $P=p_1+p_2+p_3$, $K = p_1+p_2$, $\ell = p_3-1/3 P$, $k = \frac{p_1-p_2}{2}$. The $j$ sum runs over the isospin projections. The functions $\G$ are the diquark correlations amplitudes, $\Delta_0$ and $\Delta_{\mu\nu}^1$ are the diquark propagators. Finally, the functions $\mathcal{S}$ and $\mathcal{A}_{\nu}^j$ are the quark-diquark Faddeev amplitudes. We left the colour structure implicit as it generates only an overall prefactor absorbed in the normalisation constant.

We do not tackle the computation of the PDA directly. Instead, we compute the general form of its Mellin moments, and deduce the PDA from this general expression. They are defined through:
\begin{equation}
\label{eq:MellinMomentDef}
\langle x_1^l x_2^m \rangle = \int \mathcal{D}x\, x_1^l\, x_2^m\, \da ,
\end{equation}
where, due to momentum conservation, only two indices are necessary to obtain their entire set. Putting eq. \eqref{eq:MellinMomentDef} in perspective with eq. \eqref{eq:PhiasMatrix1}, it is straightforward to realised that the Mellin moments are expectation values of local operators. Consequently, we do not need to handle Euclidean, complex valued ``light-like'' combinations of the Faddeev amplitude internal momenta. The moments are given through the projection of the Faddeev Amplitude with:
\begin{align}
  \label{eq:MomentProjection}
  \frac{i}{2} f_B p\cdot n \slashed{n} B^\uparrow \langle x_1^l x_2^m \rangle &= \int \mathcal{D}x x_1^lx_2^m \int \frac{\textrm{d}^4\ell}{(2\pi)^4}\nonumber \\
  &\times \int \frac{\textrm{d}^4k}{(2\pi)^4} \prod_{i=1}^3 \delta\left(x_i-\frac{p_i\cdot n}{P\cdot n} \right) \chi (p_1,p_2,p_3) O^\varphi_{21}O^\varphi_3
\end{align}
where the projection operators are:
\begin{equation}
  \label{eq:operators}
  O^\varphi_{21} = L^\downarrow C^\dagger \slashed{n}L^\uparrow , \quad O^\varphi_3 = \slashed{n}L^\uparrow.
\end{equation}

\section{Examples for a scalar diquark}

We exemplify the computation of the PDA based on the scalar diquark component of the Faddeev Wave Function. The leading twist projection of eq. \eqref{eq:MomentProjection} can be written in terms of $\gamma \cdot \mathcal{L}^0$ with:
\begin{eqnarray}
  \label{eq:L0Definition}
  \mathcal{L}^{0\nu} & = & \frac{1}{4} \textrm{Tr}\left[\gamma^\nu \slashed{n}L^\uparrow S(p_3) \tilde{\Gamma}^{0} \tilde{S}(p_2) L^\downarrow C^\dagger \slashed{n}  L^\uparrow S(p_1) \mathcal{S} \right]\Delta^0(K) \nonumber \\
  & = & \frac{1}{4} \textrm{Tr}\left[ S(p_3) \tilde{\Gamma}^{0} \tilde{S}(p_2) L^\downarrow C^\dagger\slashed{n} L^\uparrow \right] \textrm{Tr}\left[\gamma^\nu \slashed{n}L^\uparrow S(p_1)\mathcal{S} \right] \Delta^0(K),
\end{eqnarray}
where $S$ is the quark propagator.
The scalar diquark contribution to the leading twist PDA can therefore been split into two parts, one being the PDA of the scalar diquark itself, the second the projected quark-diquark amplitude. These two pieces are unsurprisingly convoluted through the momentum of the diquark.

A rigorous evaluation of the PDA certainly requires the computation of $\mathcal{L}^0$ using propagators and amplitudes computed within a QCD-connected, symmetry preserving framework. However, multiple example have highlighted the fact that, the use of algebraic models based on Perturbative Integral Representation (PTIR) reveals itself already insightful \cite{Chang:2014lva,Mezrag:2014tva,Mezrag:2014jka,Shi:2015esa,Chen:2016sno,Mezrag:2016hnp}. We will therefore proceed with developing such a model for the nucleon Faddeev Amplitude.

\subsection{Scalar diquark Model and Structure}

We start by modelling and evaluating the structure of the scalar diquark itself, following eq. \eqref{eq:L0Definition}. For this, we model the quark propagator and the diquark correlation vertex through:
\begin{equation}
  \label{eq:ScalarModel}
  S(p) = (-i\slashed{p}+M)\sigma_M(p^2), \quad \eta_0\Gamma(k,K)C^\dagger = i\g_5\int\textrm{d}z (1-z^2)\sigma_{\Lambda_\Gamma}(q_+)
\end{equation}
where $\sigma_M(p^2) = (p^2+M^2)^{-1}$ and $q_+ = q+(z/2) K$. We then used the methods developed in Ref. \cite{Chang:2013pq,Mezrag:2016hnp} in order to compute the Mellin moments of the Scalar diquark PDA. We therefore introduce:
\begin{equation}
  \label{eq:D0Def}
  \mathcal{D}_0^m(K^2) = \frac{1}{P\cdot n}\int\frac{\textrm{d}^4k}{(2\pi)^4}\left(\frac{k\cdot n}{P\cdot n} \right)^m\textrm{Tr}\left[ S(p_3) \tilde{\Gamma}^{0} \tilde{S}(p_2) L^\downarrow C^\dagger\slashed{n} L^\uparrow \right],
\end{equation}
use a Feynman parametrisation to rearrange the denominator and introduce a carefully thought change of variable, in such a way that we can obtain:
\begin{equation}
  \label{eq:D0Calculation}
  \mathcal{D}_0^m(K^2) = \eta_0' \left(\frac{K\cdot n}{P\cdot n}\right)^{m+1}\int  \frac{\textrm{d}v\textrm{d}u\textrm{d}\beta (1-z^2(\beta,u,v))\beta ^m}{\left[(\beta(v(\beta-2)+\beta)+u(v-\beta^2)\right]\left[\mathcal{M}^2 +K^2 \right]},
\end{equation}
with $z(\beta,u,v)=-1+2(u-\beta)/(u-v)$, $0 \le v \le \beta \le u \le 1$ and:
\begin{equation}
  \label{eq:Meff}
  \mathcal{M} = 4\frac{(1-u+v)M^2 +(u-v)\Lambda_\Gamma^2}{\left(\beta(v(\beta-2)+\beta)+u(v-\beta^2) \right)} \left(u-v \right).
\end{equation}
Eq. \eqref{eq:D0Calculation} allows us to directly extract a pointwise expression for the DA, as the expansion of a continuous function in terms of Mellin moments is unique. In the case where $M=\Lambda_\Gamma$, it is actually even possible to obtain a simple, algebraic result for the structure of the diquark:
\begin{equation}
  \label{eq:ScalarDiquarkResult}
  \eta_0''(y) \varphi_0(\hat{x}_2,\hat{x}_3) = 12y \left(1-\frac{y}{\hat{x}_2\hat{x}_3}\ln\left[1+\frac{\hat{x}_2\hat{x}_3}{y} \right] \right),
\end{equation}
with $y=M^2/K^2$, and $ \eta_0''(y)$ ensuring the PDA is normalised to 1 for every $y$. Interestingly, this result is compatible with previous continuum studies of the pion PDA \cite{Chang:2013pq,Qin:2017lcd}; namely the limit when $y\gg 1$, \emph{i.e.} $K^2 \ll \Lambda_\Gamma^2$ the PDA goes to the asymptotic one : $\varphi_0(\hat{x}_2\hat{x}_3) \rightarrow 6\hat{x}_2\hat{x}_3$. This can be understood as the correlation amplitude’s momentum-space extent is far larger than the bound-state's mass-scale, yielding an effectively scale-free system. On the other hand, when  $y\ll 1$, \emph{i.e.} $K^2 \gg \Lambda_\Gamma^2$ the system tends to look like a pointlike particle and $\varphi(\hat{x}_2,\hat{x}_3)\rightarrow 1$. One should note that the end-point behaviour of our PDA is linear, independently of $y$.

\subsection{Scalar Quark-Diquark Amplitude}
\begin{figure}[t]
  \centering
  \begin{tabular}[h]{cc}
    \includegraphics[width=0.34\textwidth]{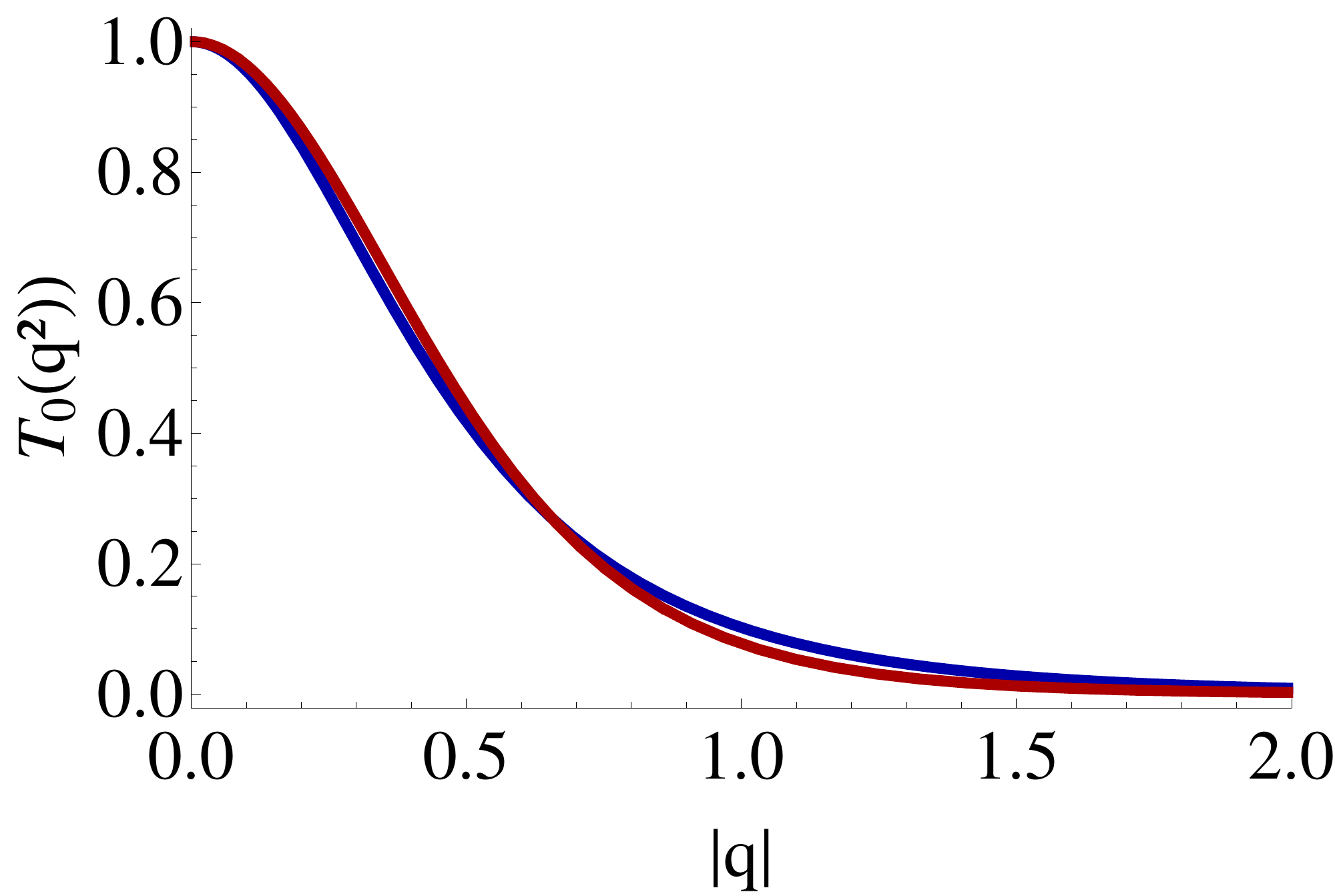} & \includegraphics[width=0.34\textwidth]{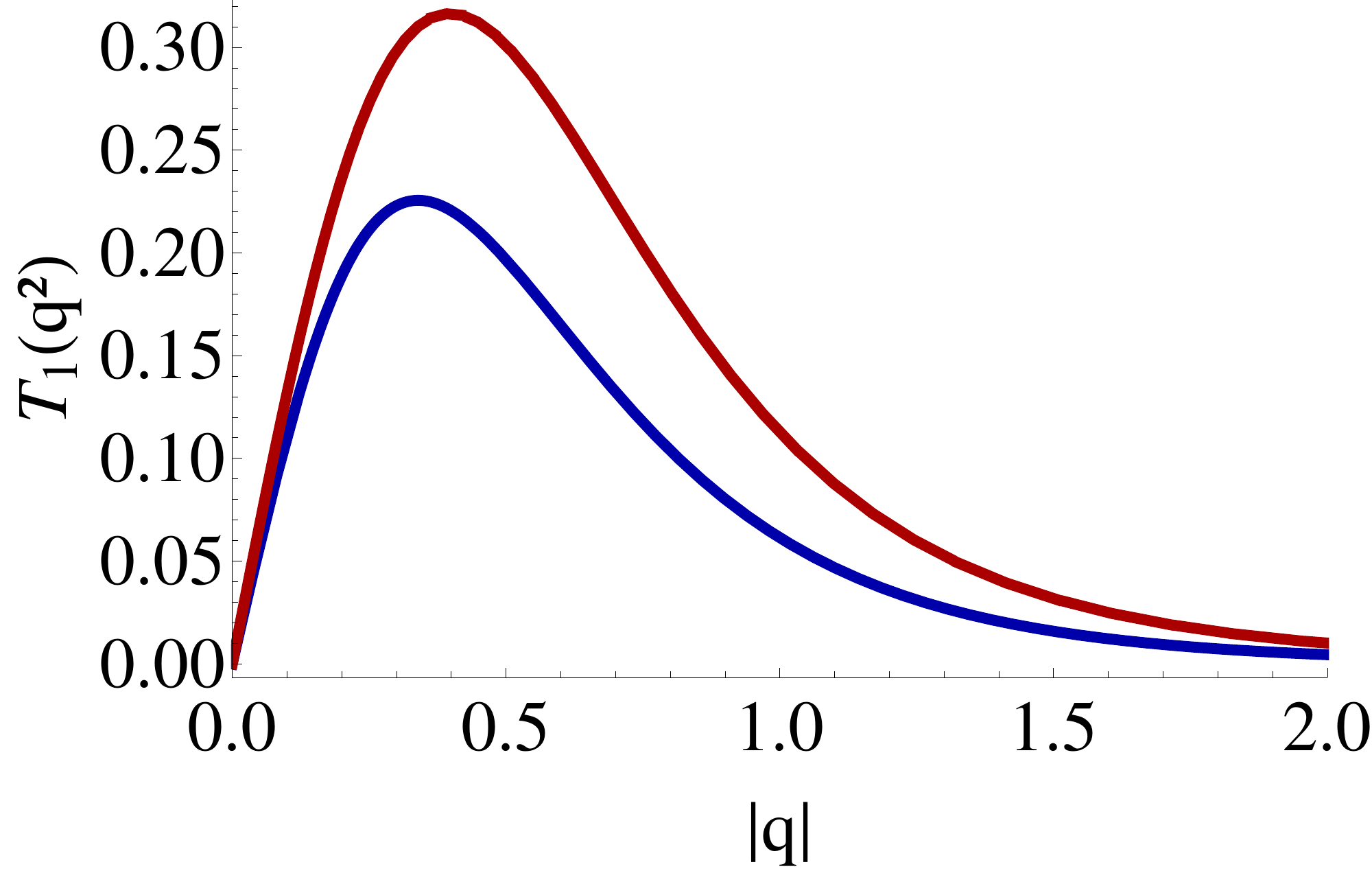} \\
    \includegraphics[width=0.34\textwidth]{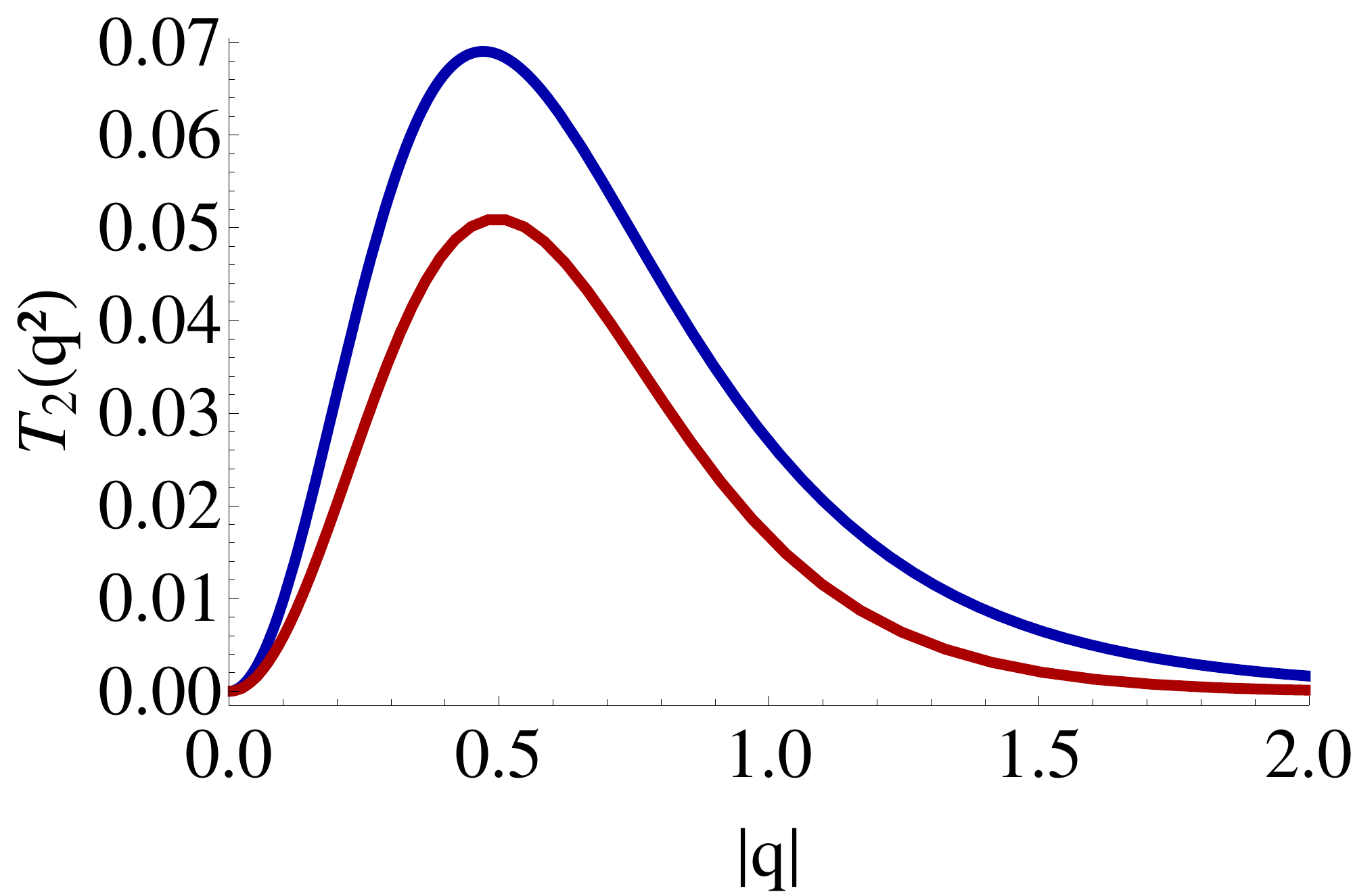} & \includegraphics[width=0.34\textwidth]{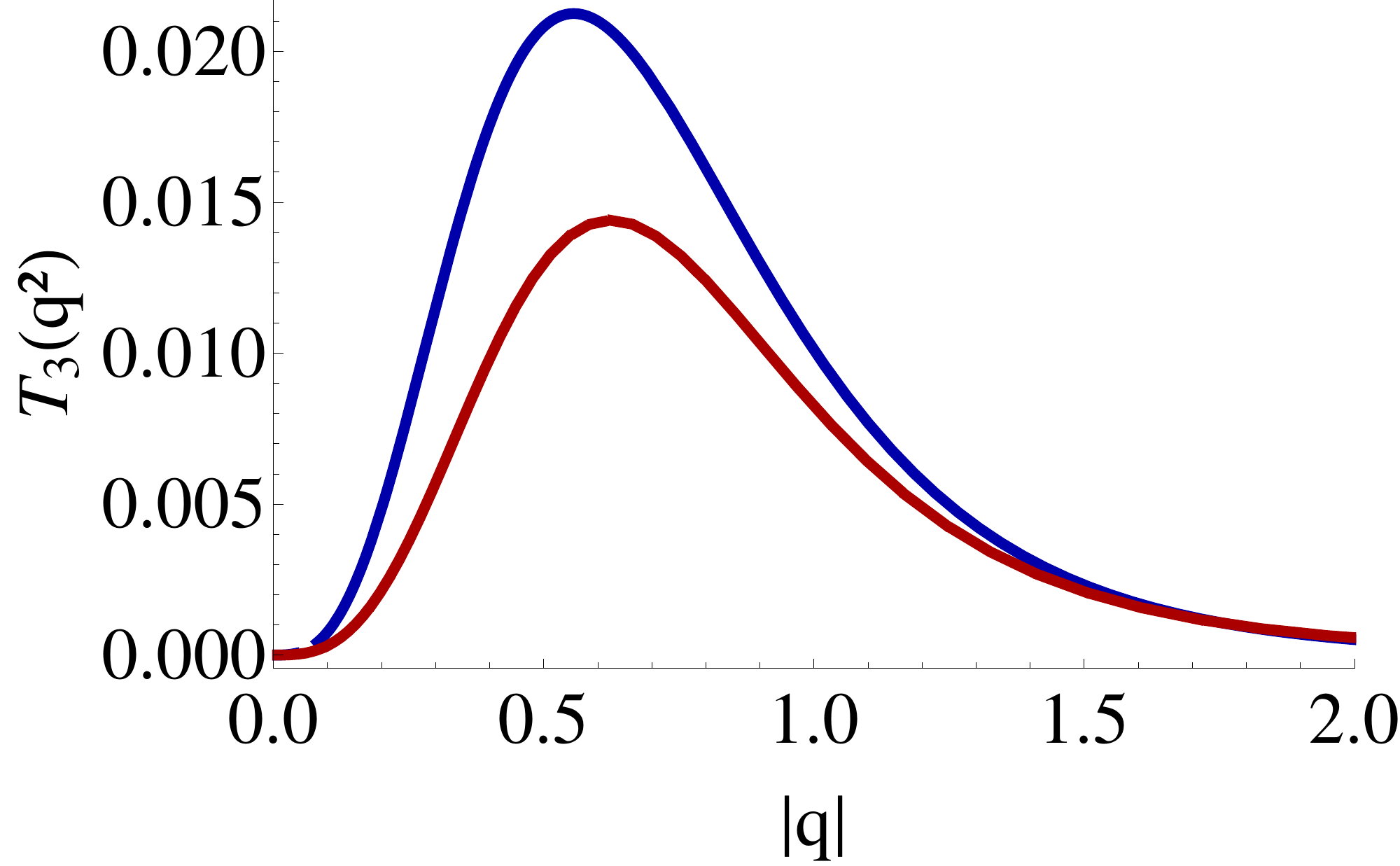}
  \end{tabular}
  \caption{Comparison of the 4 first Chebychev moments of the nucleon Faddeev Amplitude. Blue curves, results of the present model; red curves, realistic solution of the Faddeev Amplitude from Ref. \cite{Segovia:2015hra}.}
  \label{fig:RhoCompaison}
\end{figure}

\begin{figure}[b]
  \centering
  \includegraphics[width=0.4\textwidth]{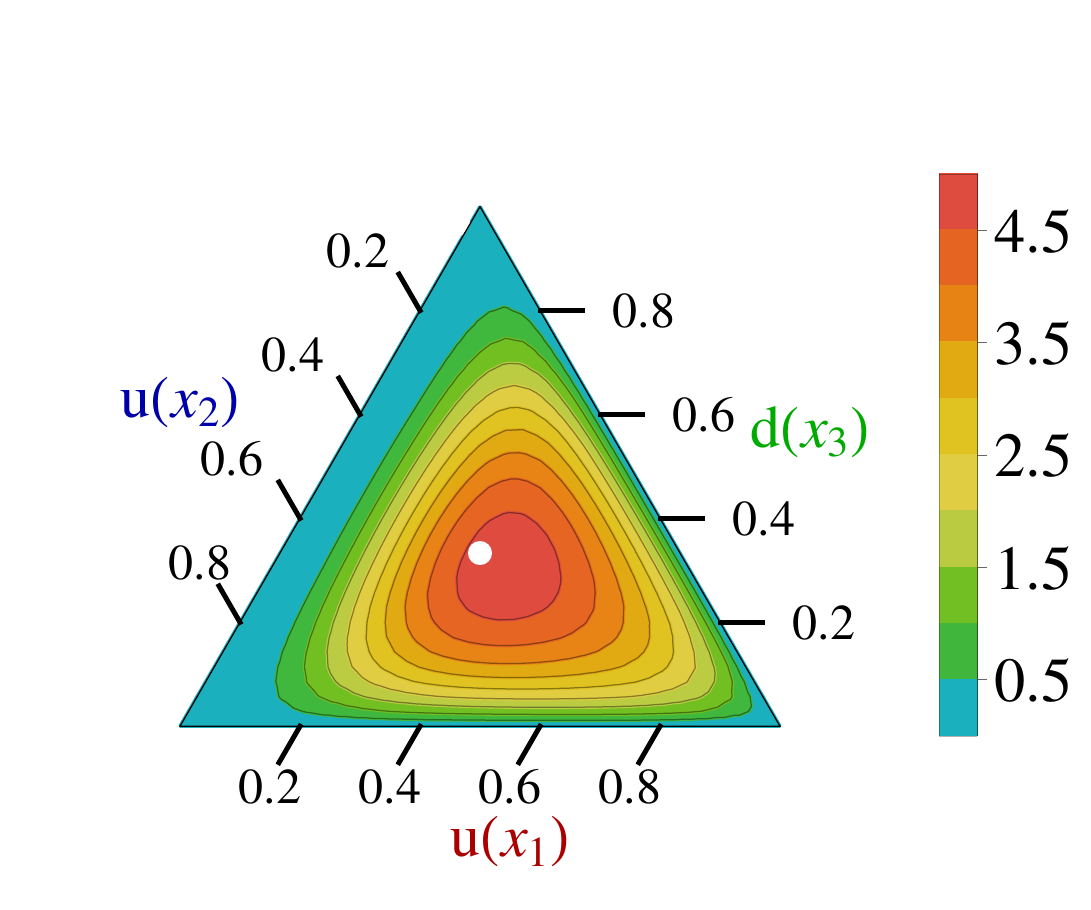} \quad \includegraphics[width=0.38\textwidth]{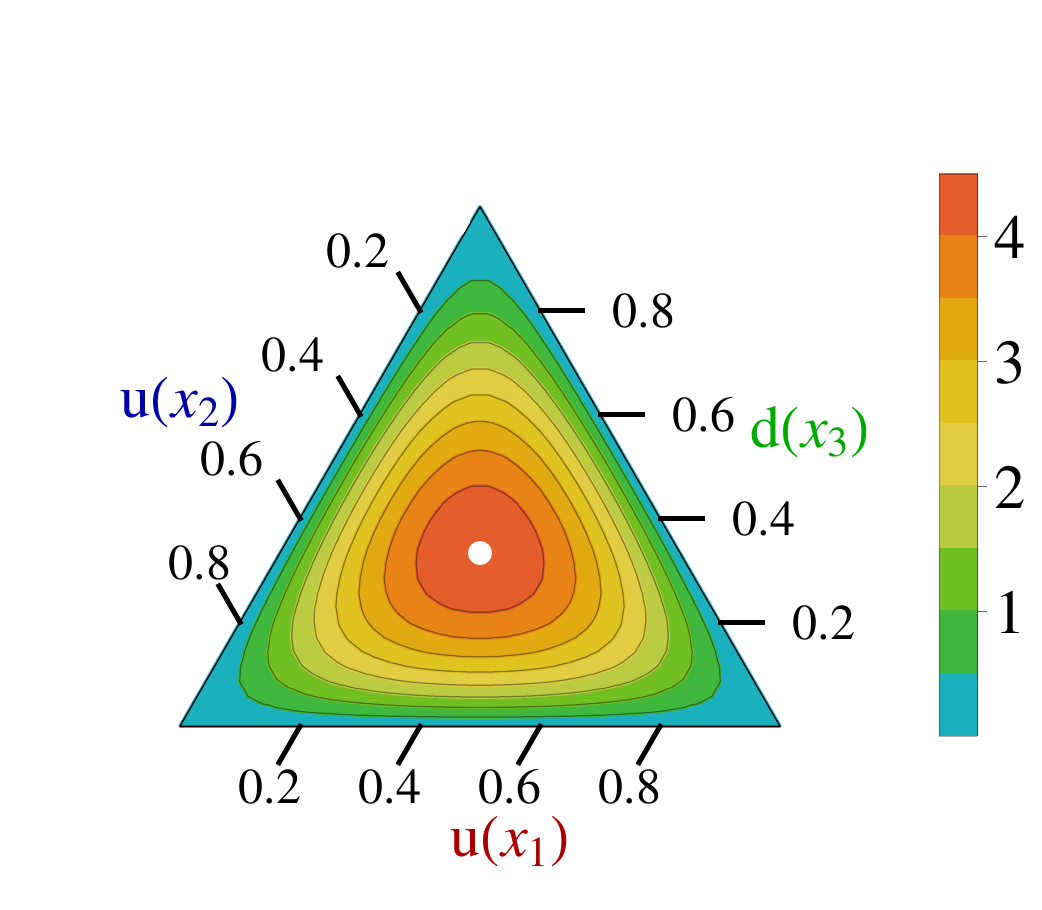} 
  \caption{Left: Result obtained for the computation of the Scalar diquark contribution using the parameters (in unit of the nucleon mass): $\{M,M_0,\Lambda_\G,\Lambda_0\}=\{2/5,9/10,3/5,6/5\}$; Right: asymptotic nucleon PDA. }
  \label{fig:Results}
\end{figure}

In order to obtain the scalar diquark contribution to the nucleon PDA, it is necessary to perform the convolution of our scalar diquark structure with the quark-diquark Faddeev wave function. The latter is modelled using an effective diquark propagator and a Faddeev amplitude also computed thanks to PTIR:
\begin{equation}
  \label{eq:FaddeevModel}
  \Delta_0(K^2) = \sigma_{M_0}(K^2), \quad \mathcal{S}(\ell,P) = i\eta \int \textrm{d}z (1-z^2)\rho(z)\sigma^3_{\Lambda_0}(\ell_-^2), 
\end{equation}
with $\ell_- = \ell-(1+3z)/6P$. In the purpose of determining a realistic weight, we expand the $\rho$ function on the $3/2$-Gegenbauer polynomial basis and we tune the coefficients in order to reproduce the first Chebychev moments coming from realistic numerical solution of the Faddeev equation \cite{Segovia:2014aza,Segovia:2015hra}. 
The results, shown in Fig.\ref{fig:RhoCompaison}, are in fair but not perfect agreement with the realistic computations. From that point, we apply the same computing strategy than before in order to integrate the system over $\ell$. The results are shown on Fig.\ref{fig:Results}.
The nucleon PDA appears to be skewed with respect to the asymptotic one, emphasising that the bystander quark is more likely to carry the baryon lightfront momentum than the two quarks forming the strong diquark correlation. 

\section{Conclusion}
We present here a new step toward realistic models of the baryon PDA, by improving our previous work and presenting the preliminary results we obtain in the scalar case. The results are qualitatively identical to the previous ones, the scalar distribution is skewed, emphasising the bystander quark, but quantitatively different as they are now less skewed. We now look at extending our improved description of the Nakanishi weight functions to the Axial-vector diquark contributions, and beyond to the first radial excitation of the nucleon, namely the Roper resonance.

\subsection*{Acknowledgement}
We are grateful for insightful comments from F. Gao, V. Mokeev, H. Moutarde, S.-X. Qin, J. Rodr\'iguez-Quintero, G. Salm\`e and S.-S. Xu. Work supported by: European Union’s Horizon 2020 research and innovation programme  under  the  Marie  Sklodowska-Curie Grant Agreement No. 665919; Spanish MINECO’s Juan de la Cierva-Incorporaci on programme, Grant Agreement No. IJCI-2016-30028; and MINECO Contract Nos. FPA2014-55613-P, FPA2017-86989-P and SEV-2016-0588; the Chinese Government's Thousand Talents Plan for Young Professionals; the Chinese Ministry of Education, under the International Distinguished Professor programme; and U.S. Department of Energy, Office of Science, Office of Nuclear Physics, under contract no. DE-AC02-06CH11357.

%
\bibliographystyle{bibtex/spphys}
\bibliography{Bibliography}

\end{document}